\documentclass{statsoc}

\usepackage{amsmath,amssymb,natbib,subfigure,graphicx,setspace}
\usepackage{placeins}
\usepackage[text={6in,9in},right=1in]{geometry}

\newcommand{\bftheta}{\hbox{\boldmath$\theta$}}
\newcommand{\bfL}{\hbox{\boldmath$L$}}
\newcommand{\bfX}{\hbox{\boldmath$X$}}
\newcommand{\bfv}{\hbox{\boldmath$v$}}
\newcommand{\bfu}{\hbox{\boldmath$u$}}

\title[Poisson Mixtures]{Experience Rating with Poisson Mixtures}

\author[]{G. O. Brown\footnote{Email:gob20@statslab.cam.ac.uk}} 
\address{Statistical Laboratory, Centre for Mathematical Sciences, 
  Cambridge CB3 0WB, UK}
\author[]{S. P. Brooks}
\address{Statistical Laboratory, Centre for Mathematical Sciences, 
  Cambridge CB3 0WB, UK}
\author[]{W. S.  Buckley}
\address{CBA, Florida International University, Miami, FL 33199, USA}

\begin{document}
\begin{abstract}
  We present a mixture Poisson model for claims counts in which the
  number of components in the mixture are estimated by reversible jump
  MCMC methods.
\end{abstract}


\section{Introduction}
In this paper we consider a mixed Poisson model for count data arising
in Group Life insurance. We present a Bayesian formulation to
determine the number of groups in an insurance portfolio consisting of
claim numbers or deaths. We take a non--parametric Bayesian approach
to modelling this mixture distribution using a Dirichlet process prior
and use reversible jump Markov chain Monte Carlo to estimate the
number of components in the mixture. The physical interpretation of
the model in this case is that the heterogeneity is assumed to be
drawn from one of a finite number of possible groups, in proportion
which will be estimated.

\section{A Credibility Model for Heterogeneity}\label{sec:credmodelforhet}
\begin{figure}[p]
  \centering
  \includegraphics[width=0.8\textwidth]{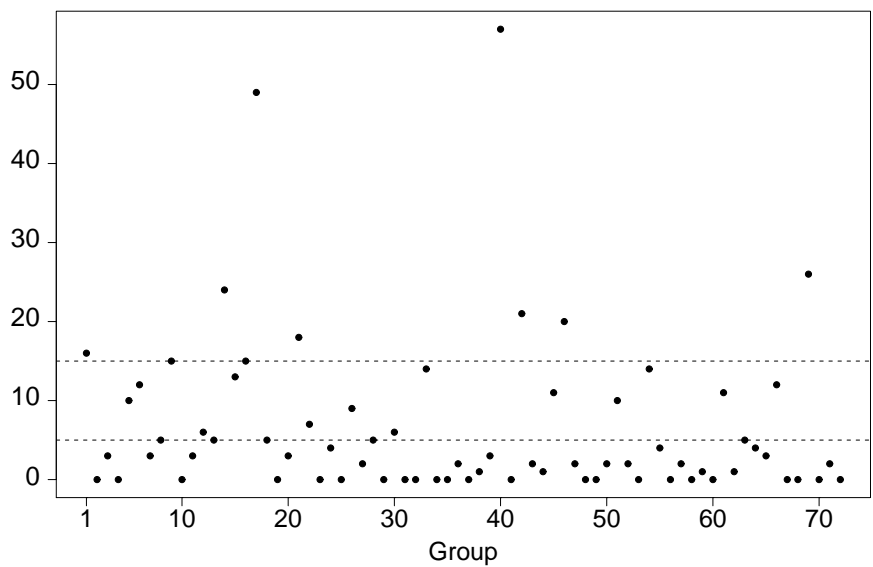}
  \caption{Plot of the number of observed claims for each Group.}
  \label{fig:heterodataabsplot}
\end{figure}

\begin{figure}[p]
  \centering
  \includegraphics[width=0.8\textwidth]{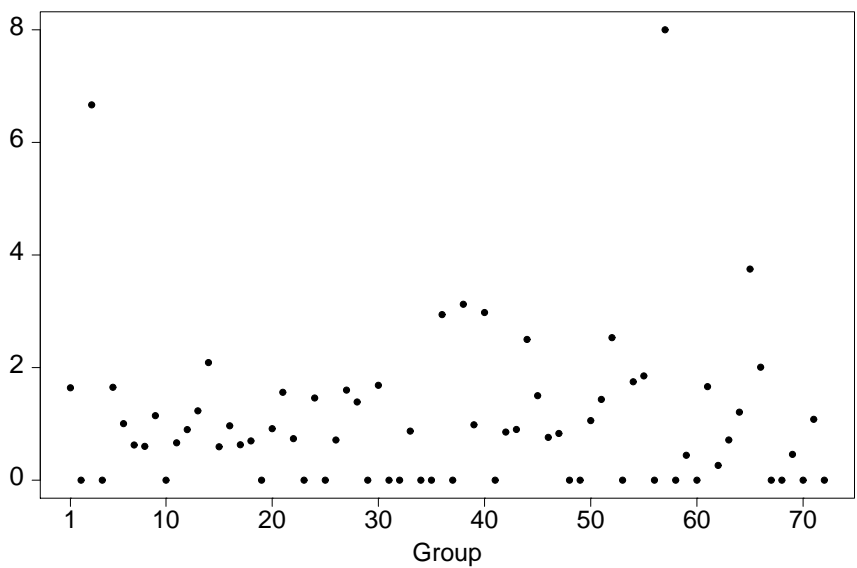}
  \caption{Plot of the number of observed claims per unit exposure.}
  \label{fig:heterodataplot}
\end{figure}

The data arise from 1125 groups insured through the whole or parts of
the period 1982--1985 by a major Norwegian insurance company. There
are $n=72$ classes distinguished by occupation category. The $i^{th}$
class has risk exposure $E_i$, and observed number of deaths, $D_i$.
The data are also analysed in \citet{haastrup2000} and
\citet{norberg1989}.  Let $D_1, \ldots, D_n$ denote the number of
observed deaths in each insured group.  Associated with each group is
the exposure, denote $E_1,\ldots,E_n$, respectively, which is a
measure of the propensity of that group to produce claims/deaths. Let
$D^n$ denote the collection of all deaths for each group, where
\begin{equation*}
  D^n = \{ D_1, \ldots, D_n\}.
\end{equation*}
Similarly, let $E^n$ denote the collection of all exposures for the
group
\begin{equation*}
  E^n =\{ E_1, \ldots, E_n\}.
\end{equation*}
Figure~\ref{fig:heterodataabsplot} shows a plot of the claim number
for each group while Figure~\ref{fig:heterodataplot} shows the claim
numbers normalized by their corresponding exposures.

The heterogeneity model is used to model differences in each of the
$n$ groups.  For each group, the exposures are recorded and the
resulting number of deaths or claims are then recorded for each
group. At the first level, the number of claims for each group is
assumed to follow a Poisson distribution with parameter
$\lambda_iE_i$, for $i=1, \ldots, n$.  Thus
\begin{equation*}
  D_i \sim \mathcal{P}oisson(\lambda_i E_i),\quad i=1, \ldots, n.
\end{equation*}
We take a fully Bayesian approach and assume that the $\lambda_i$ are
IID and follow a Gamma distribution with parameters $\alpha$ and
$\beta$; that is,
\begin{equation*}
  \lambda_i \sim\mathcal{G}amma(\alpha, \beta),
\end{equation*}
where $\alpha$ and $\beta$ are also assumed to be unknown.

The advantage of using such mixed distributions is that it
allows for extra variation in the number of occurrences since
\begin{align*}
  \mathbb{E}(D) &=
  \mathbb{E}(\mathbb{E}(D| \lambda ) ) = \mathbb{E}(E\lambda ) 
  = E\alpha / \beta \\
  \intertext{and}
  \mathbb{V}ar (D) & = \mathbb{E}(\mathbb{V}ar(D|\lambda) ) +
  \mathbb{V}ar(\mathbb{E}(D | \lambda )) \\
  &=  E\alpha /\beta + E^2 \alpha / \beta^2 
 > \mathbb{E}(D).
\end{align*}

\section{Extending the Basic Model}
In an analysis based on the model described in
Section~\ref{sec:credmodelforhet}, \citet{haastrup2000} assumes that
each group has its own unique heterogeneity parameter $\lambda$, drawn
from some distribution.  \citet{haastrup2000} assumes that each class
$i$ has a unique heterogeneity parameter, denoted $\lambda_i$, and
that the number of deaths $D_i$ in this class follows a Poisson
distribution with mean $\lambda_iE_i$.  The classes are assumed to be
mutually independent, given the heterogeneity parameters $\lambda_1$,
$\lambda_2$, \ldots, $\lambda_n$.  Furthermore, he assumes that this
distribution is identical for each group.  In practice, large values
of $E_i$ will account for large values of $D_i$, which will lead to
similar values of $\lambda_i$ for each $i$.

In our analysis, we propose a mixture model formulation. We assume
that $D_i$, given $\lambda_j$, has a Poisson distribution with mean
$\lambda_j E_i$.  We take a non--parametric Bayesian approach to
modelling this mixture distribution using a Dirichlet process prior,
and use reversible jump Markov chain Monte Carlo to estimate the
number of components in the mixture.  In this case, the physical
interpretation of the model is that the heterogeneity is assumed to be
drawn from one of $k$ possible groups, in proportions $w_1$, \ldots,
$w_k$.


\section{Mixture Formulation}\label{sec:mixformulation}
The method we describe is essentially a classification problem where
we assume that each observed $D_i$ comes from any one of $k$
components, where each component has a Poisson distribution. Thus
\begin{equation*}
  D_i|\lambda_j\sim\mathcal{P}oisson(E_i\lambda_j)
  \quad j=1,\ldots, k ; i=1, \ldots, n .
\end{equation*}
More general forms of the mixture Poisson model with covariates are
discussed in \citet{green2002:richardson}. Mixture models for grouped
claim numbers are considered by \citet{tremblay1992} and
\citet{walhin1999,walhin2000}.  \citet{dellaportas1997} considers
count data in Finance using split/merge moves, while
\citet{viallefont2002} provides a more general discussion of mixtures
of Poisson distributions, using both split/merge moves and birth/death
moves. Other methods for determining the number of components in a
mixture are discussed by \citet{mclachlan2000},
\citet{phillips1996:mcmcip}, \citet{carlin1995}, and
\citet{stephens2000} who use Markov chains to model jointly the number
of components and component values. The advantage of the Bayesian
formulation is that we can place posterior probabilities on the order
of the model.

\subsection{The Likelihood Function}
Throughout our discussion, $n$ will denote the number of data points and $k$
will denote the number of components in the mixture formulation. For a finite
mixture model, the observed likelihood function is
\begin{equation}\label{eq:mixobservedlik}
  \bfL(D^n | \lambda, w, E^n) = \prod_{i=1}^n
  \sum_{j=1}^k w_j f_j(D_i |\lambda_j, E_i),
\end{equation}
where the weights are non-negative and $\sum_{j=1}^k w_j =1$.  Even for
moderate values of $n$ and $k$, this takes a long time to evaluate since there
are $k^n$ terms when the inner sums are expanded \citep{casella2000}.
Another form of the likelihood function will be derived shortly. Classical
estimation procedures for mixture models are described by
\citet{titterington1990} and \citet{mclachlan2000}.

Let $z_i$ be a categorical random variable taking values in $\{ 1,\ldots,
k\}$ with probabilities $w_1, \ldots, w_k$, respectively, so that
\begin{equation*}
  p(z_i = j | w) = w_j.
\end{equation*}
Suppose that the conditional distribution of $D_i$, given $z_i = j$, is
$Poisson(\lambda_j)$, $j=1,\ldots,k$.  Let $f_j(\cdot)$ denote a Poisson
density with parameter $\lambda_j$. Then the unconditional density of $D_i$
is given by $f(D_i)$, where
\begin{align}
  f(D_i) &= \sum_{j=1}^k w_j f_j(D_i|\lambda, E^n) ,\nonumber \\
\intertext{since}
  f(D_i) &= \sum_{j=1}^k f_j (D_i | z_i = j, \lambda, E^n)
  p(z_i = j ) \label{eq:fofNi}\\
   &= \sum_{j=1}^k w_j f_j(D_i|\lambda_j E_i). \nonumber
\end{align}
With every pair $(D_i, E_i)$, we associate a latent variable $z_i$, which is an
indicator variable that indicates which component of the mixture is
associated with $(D_i, E_i)$. We have, $z_i=j$ if the $i^{th}$ data point,
$(D_i, E_i)$, comes from the $j^{th}$ component of the mixture. Thus, for each
$i$, we have
\begin{equation*}
  z_i | w\sim \mathcal{M}(1; w_1, \ldots, w_k)
\end{equation*}
and
\begin{equation*}
  D_i | z_i \sim \mathcal{P} (\lambda_{z_i} E_i)  .
\end{equation*}
By incorporating the indicator variables $z$, the complete data likelihood
is then
\begin{align}\label{eq:mixcompletelik}
  \bfL(D^n | z, \lambda, E^n) &=
  \prod_{i=1}^n f(D_i | \lambda_{z_i}, E_i) \nonumber \\
  &= \prod_{j=1}^k \prod_{\{i:z_i=j \}} f(D_i | \lambda_j, E_i).
\end{align}
At times, especially for the fixed $k$ case described below, it is more
convenient to work with \eqref{eq:mixcompletelik} as it involves
multiplications only, rather than additions and multiplications, as in
\eqref{eq:mixobservedlik}.

The convenience of using the missing data formulation is that the posterior
conditional distribution of the model parameters would be standard
distributions. Also the augmented variables $z$ allows us to see what
component of the mixture the data points are assigned.


\subsection{Gibbs Updates for Fixed $k$}\label{sec:fixedk}
We consider a mixture of Poissons where, conditional on there being $k$
components in the mixture, we have
\begin{equation*}
  D_i\sim \sum_{j=1}^k w_j f(\cdot| \lambda_j, E_i).
\end{equation*}
The weights $w_j$ sum to one, and are non--negative, so that
\begin{equation}\label{eq:sumtoone}
  \sum_{j=1}^k w_j = 1, \text{ and } w_j \geq 0.
\end{equation}
\begin{equation*}
  f(D_i| \lambda, z_i =j, E_i) \sim \text{Poisson}(\lambda_jE_i)
  \text{ with } P(z_i=j) = w_j
\end{equation*}
and
\begin{equation*}
  w\sim \mathcal{D}(\delta_1, \ldots, \delta_k)
\end{equation*}
follows a Dirichlet distribution.
We also make the additional assumption that the $\delta_j$'s are equal to $1$
so that $p(w)$ is a uniform distribution on the space described by
\eqref{eq:sumtoone}. For the Poisson parameters $\lambda_j$, we take Gamma
priors, so that
\begin{equation*}
  \lambda_j\sim\text{Gamma}(a,b) \quad j=1,\ldots, k
\end{equation*}
with the ordering constraint
\begin{equation}\label{eq:ordering}
  \lambda_1 < \lambda_2 < \ldots < \lambda_k,
\end{equation}
to ensure that the components are identifiable.
The ordering constraint is not necessary for the Monte Carlo algorithm to
work.  However it does avoid the problem of label switching, since otherwise,
any permutation of the indices $\{1,\ldots, k\}$ will result in the same
posterior distribution.

Note that because we have the ordering constraint in
Equation~\eqref{eq:ordering}, the joint density of the collective $\lambda$ is
\begin{equation*}
  p(\lambda|\alpha, \beta, k) =
  k! p(\lambda_1|\alpha,\beta) \cdots p(\lambda_k|\alpha,\beta)
  I_{  \lambda_1 < \lambda_2 < \ldots < \lambda_k}(\lambda).
\end{equation*}
When $k$ is fixed and known, the factorial term $k!$ does not affect the MCMC
algorithm since it can be absorbed into the normalising constant. However, in
the variable $k$ case, it must be noted, since it is a factor in the reversible
jump acceptance probability.  The joint density of all unknowns is
\begin{equation} \label{eq:heterposterior}
  \pi(w, \lambda, z | D^n) \propto
  p(w|\delta) p(z|w) p(\lambda |\alpha, \beta) \bfL (D^n |\lambda, z, E^n).
\end{equation}
With the missing data formulation, the likelihood term $\bfL(D^n|\lambda, z,
E^n)$ can be written
\begin{equation*}
  \bfL(D^n |\lambda, z, E)=  \prod_{i=1}^n \left(
    \frac{e^{-\lambda_{z_i}E_i}(\lambda_{z_i} E_i)^{D_i}}{D_i!}
  \right),
\end{equation*}
and
\begin{equation*}
  p(z|w) =  \prod_{j=1}^k w_j^{n_j},
\end{equation*}
where $n_j = \#\{i | z_i = j\}$, is the number of observations allocated to
component $j$.
The prior distributions are
\begin{equation*}
  p(w|\delta) =
  \frac{\Gamma(\sum_{j=1}^k \delta_j)}{\prod_{j=1}^k \Gamma(\delta_j)}
  \prod_{j=1}^k w_j^{\delta_j-1}
\end{equation*}
\begin{equation*}
  p(\lambda_i| \alpha, \beta) = \frac{\beta^{\alpha} }{\Gamma(\alpha)}
  \lambda_i^{\alpha-1} e^{-\beta\lambda_i} .
\end{equation*}

Using Bayes' theorem, we have the following posterior conditional
distributions:
\begin{equation*}
  \pi(\lambda_j) \propto
  \lambda_j^{\alpha-1} e^{-\beta\lambda_j} \times
  \lambda_j^{\sum_{i|z_i=j} D_i}
  e^{-\lambda_j\sum_{i|z_i=j}E_i}I_{(\lambda_{j-1},\lambda_{j+1})}(\lambda_j)
\end{equation*}
and
\begin{equation*}
  \pi(w | \delta, z) \propto p(w| \delta) p(z | w)
\end{equation*}
so that,
\begin{equation*}
  w \sim \mathcal{D}(\delta_1+n_1, \ldots, \delta_k+n_k),
\end{equation*}
where $n_j = \#\{i|z_i=j \}$. For $z$, we update the allocations using
\begin{equation*}
  P(z_i = j) \propto w_j f(D_i| \lambda_j, E_i)
  \quad i=1,\ldots,n; \quad j=1,\ldots,k,
\end{equation*}
so that,
\begin{equation}\label{eq:allocconditional}
  p(z_i = j) = \frac{w_j f(D_i | \lambda_j, E_i)}
  {\sum_{j=1}^k w_j f(D_i | \lambda_j, E_i)}.
\end{equation}
This follows from Equation~\eqref{eq:fofNi}. The Gibbs algorithm for fixed
$k$ is then \citep{robert1999}
\begin{description}
  \item[Step 1:] Simulate $z_i$ from
    \begin{equation*}
      p(z_i = j) \propto w_j f(D_i | \lambda_j, E_i)\text{ for $j=1,\ldots, k$}
    \end{equation*}
    and compute $n_j$, $n_j\bar{D}_j$, $n_j\bar{E}_j$ from
    \begin{equation*}
      n_j = \sum_{i|z_i=j} (1) \quad
      n_j\bar{D}_j = \sum_{i|z_i=j} D_i\quad
      n_j\bar{E}_j = \sum_{i|z_i=j} E_i
    \end{equation*}
  \item[Step 2:] Simulate
    \begin{equation*}
    \lambda_j \sim
    \mathcal{G}amma( \alpha + n_j\bar{D}_j, \beta + n_j \bar{E}_j )
    \mathbb{I}_{(\lambda_{j-1}, \lambda_{j+1})}(\lambda_j)
    \text{ for $j=1,\ldots,k$}
    \end{equation*}
  \item[Step 3:] Simulate
    \begin{equation*}
      w \sim \mathcal{D}(\delta + n_1, \ldots, \delta + n_k) .
    \end{equation*}
\end{description}

\subsection{Reversible Jump MCMC}
The Reversible jump algorithm is an extension of the
Metropolis--Hastings algorithm.  We assume there is a countable
collection of candidate models, indexed by $M\in\mathcal{M}=\{M_1$,
$M_2$,$\ldots$ , $M_k\}$. We further assume that for each model $M_i$,
there exists an unknown parameter vector $\bftheta_i \in
\mathbb{R}^{n_i}$ where $n_i$, the dimension of the parameter vector,
can vary with $i$. Typically we are interested in finding which models
have the greatest posterior probabilities and also estimates of the
parameters. Thus the unknowns in this modelling scenario will include
the model index $M_i$ as well as the parameter vector $\bftheta_i
$. We assume that the models and corresponding parameter vectors have
a joint density $\pi(M_i,\bftheta_i )$.  The reversible jump algorithm
constructs a reversible Markov chain on the state space $\mathcal{M}
\times \bigcup_{M_i\in\mathcal{M}} \mathbb{R}^{n_i}$ which has $\pi$
as its stationary distribution \citep{green1995}.  In many instances,
and in particular for Bayesian problems this joint distribution is of
the form
\begin{equation*}
  \pi(M_i, \bftheta_i) = \pi(M_i, \bftheta_i\vert
  \bfX)\propto
  \bfL(\bfX\vert M_i, \bftheta_i) \; 
  p(M_i, \bftheta_i ) ,
\end{equation*}
where the prior on $(M_i,\bftheta_i)$ is often of the form
\begin{equation*}
  p(M_i, \bftheta_i) = p(\bftheta_i\vert M_i) \;p(M_i)
\end{equation*}
with $p(M_i)$ being the density of some counting distribution.

Suppose now that we are at model $M_i$ and a move to model $M_j$ is proposed
with probability $r_{ij}$. The corresponding move from
$\bftheta_i$ to $\bftheta_j$ is achieved by using a
deterministic transformation $h_{ij}$, such that
\begin{equation}\label{eq:revjumpmapping}
  (\bftheta_j , \bfv) = h_{ij}(\bftheta_i, \bfu ),
\end{equation}
where $\bfu$ and $\bfv$ are random variables introduced to ensure dimension
matching necessary for reversibility. To ensure dimension matching we must
have
\begin{equation*}
  \dim(\bftheta_j)+\dim(\bfv)=\dim(\bftheta_i)+\dim(\bfu). 
\end{equation*}
For discussions about possible choices for the function $h_{ij}$ we refer the
reader to \citet{green1995}, and \citet{brooks2003}.  Let 
\begin{equation}\label{eq:acceptratio}
  A(\bftheta_i, \bftheta_j ) = 
  \frac{\pi(M_j, \bftheta_j)}{\pi(M_i, \bftheta_i )}  
  \frac{q(\bfv)}{q(\bfu)} 
  \frac{r_{ji}}{r_{ij}}\hskip .15cm
  \biggl\lvert\frac{\partial h_{ij}(\bftheta_i, \bfu)}
  {\partial (\bftheta_i,\bfu)} \biggr\rvert
\end{equation}
then the acceptance probability for a proposed move from model $(M_i,
\bftheta_i)$ to model $(M_j, \bftheta_j)$ is
\begin{equation*}
  \min\left\{1, A(\bftheta_i, \bftheta_j )  \right\}
\end{equation*} 
where $q(\bfu)$ and $q(\bfv)$ are the respective proposal densities for
$\bfu$ and $\bfv$, and $\lvert \partial h_{ij}(\bftheta_i,\bfu) / \partial
(\bftheta_i,\bfu) \rvert$ is the Jacobian of the transformation induced by
$h_{ij}$. \citet{green1995} shows that the algorithm with acceptance
probability given above simulates a Markov chain which is reversible and
follows from the detailed balance equation
\begin{equation*}
  \pi(M_i, \bftheta_i) q(\bfu) r_{ij} = 
  \pi(M_j, \bftheta_j) q(\bfv) r_{ji} \hskip .1cm \biggl\lvert
  \frac{\partial h_{ij}(\bftheta_i,\bfu )} 
  {\partial (\bftheta_i, \bfu)}
  \biggr\rvert .
\end{equation*}
Detailed balance is necessary to ensure reversibility and is a sufficient
condition for the existence of a unique stationary distribution.
For the reverse move from model $M_j$ to model $M_i$ it is easy to see that
the transformation used is $(\bftheta_i, \bfu) =
h_{ij}^{-1}(\bftheta_j, \bfv)$ and the acceptance probability for
such a move is
\begin{equation*}
  \min\left\{1, 
    \frac{\pi(M_i, \bftheta_i)}{\pi(M_j, \bftheta_j)}  
    \frac{q(\bfu)}{q(\bfv)} 
    \frac{r_{ij}}{r_{ji}}\hskip .15cm
    \biggl\lvert\frac{\partial h_{ij}(\bftheta_i,\bfu)}
    {\partial (\bftheta_i,\bfu)} \biggr\rvert^{-1} \right\} 
  =\min\left\{1, A(\bftheta_i, \bftheta_j)^{-1} \right\}.
\end{equation*} 
For inference regarding which model has the greater posterior probability we
can base our analysis on a realisation of the Markov chain constructed above.
The marginal posterior probability of model $M_i$ 
\begin{equation*}
  \pi(M_i\vert \bfX) = \frac{p(M_i) f(\bfX\vert M_i)}
  {\sum_{M_j\in\mathcal{M}} p(M_j) f(\bfX\vert M_j) },
\end{equation*}
where 
\begin{equation*}
  f(\bfX\vert M_i) =\int \bfL(\bfX\vert 
  M_i,\bftheta_i)
  p(\bftheta_i|M_i)\, d\,\bftheta_i
\end{equation*} 
is the marginal density of the data after integrating over the unknown
parameters $\bftheta$. In practice we estimate $\pi(M_i| \bfX)$ by counting
the number of times the Markov chain visits model $M_i$ in a single long run
after reaching stationarity.  These between model moves described in this
section are also augmented with within model Gibbs updates as given in
Section~\ref{sec:fixedk} to update model parameters.

\section{Reversible Jump Model Selection}\label{sec:heterrevjump}
To update the model order and thereby increase or decrease the number of
components in the mixture, we use a combination of birth/death and split/merge
moves as described below.  We assume a uniform prior on the number of
components $k$, so that
\begin{equation*}
  k \sim U\{1,\ldots, k_{max} \},
\end{equation*}
where $k_{max}$ is chosen to allow the algorithm to explore all feasible
models. We set $k_{max}=72$, the number of groups, as under our hypothesis,
this is the maximum number of components in the mixture.  $k=k_{max}$ only
when the groups are all distinct. Setting $k_{max}=72$ will allow
for direct comparison of the empirical Bayesian and the mixture model
approach.

Introducing a prior on the number of components $k$, we extend the joint
density \eqref{eq:heterposterior} of all parameters. Thus, now
\begin{equation}\label{eq:revjumpall}
  \pi(k, w, z, \lambda| D^n) \propto
  p(k) p(w | \delta, k) p(\lambda | \alpha,\beta, k) p(z| k)
  \bfL(D^n|\lambda, z ).
\end{equation}
Note that the densities of the other model parameters now depend on $k$.  In
Sections~\ref{sec:splitmerge} and \ref{sec:birthdeath}, we describe in detail
two algorithms which are used to simulate from this density. These algorithms
are then combined with the fixed $k$ updates of Section~\ref{sec:fixedk} to
simulate from the density in Equation~\eqref{eq:revjumpall}. Modelling
mixtures with and without the Dirichlet process prior is considered by
\citet{green2001:scand}, who also considers the case of an unknown number of
components. Alternatives to the reversible jump algorithm in this context do
exist, see for example \citet{dellaportas2001}, who develop a semi--parametric
sample based method to approximate a mixing density $g(\bftheta)$ based on
the method of moments.

\subsection{Split and Merge Moves}\label{sec:splitmerge}
Note that the joint density in Equation~\eqref{eq:revjumpall} now depends on
$k$. We use the split/merge method of \citet{dellaportas1997} and
\citet{viallefont2002}. Suppose we are at a configuration with $k$
components, let
\begin{equation*}
  \bftheta_k = \{ (\lambda_1, w_1), \ldots, (\lambda_k, w_k) \},
\end{equation*}
and suppose a move to increase the number of components is proposed. We
select uniformly one of the current $k$ components to be split.  Suppose the
$j^{th}$ component, $(\lambda_j, w_j)$, is selected to be split into two
components $(\lambda_{j_1}, w_{j_1})$ and $(\lambda_{j_2}, w_{j_2})$ such
that $j_1=j$ and $j_2=j+1$, the components originally numbered $j+1, \ldots,
k$ are then renumbered $j+2, \ldots, k+1$.  The split is also designed so
that the first two moments of the split component remains the same as the
original component. Thus, we simulate $u_1$ and $u_2$ from densities defined
on the interval $[0,1]$. Usually, we use Beta densities and set
\begin{eqnarray*}
    w_{j_1}       &= &w_j u_1, \\
    w_{j_2}       &= &w_j (1-u_1), \\
    \lambda_{j_1} &= &\lambda_j u_2, \\
    \lambda_{j_2} &= &\lambda_j (1-u_1 u_2)/ (1-u_1).
\end{eqnarray*}
Other choices for splitting and merging components are described in
\citet{viallefont2002}. The proposed parameter is then
\begin{equation*}
  \bftheta_{k+1} = \{
  (\lambda_1, w_1), \ldots, (\lambda_{j-1}, w_{j-1}),
  (\lambda_{j_1}, w_{j_1}),   (\lambda_{j_2}, w_{j_2}),
  (\lambda_{j+1}, w_{j+1}), \ldots, (\lambda_k, w_k)
  \} .
\end{equation*}

If the ordering constraint in Equation~\eqref{eq:ordering} is not satisfied
then the move is rejected immediately, as the reverse move in which we merge
two adjacent components would not be possible. We can compute the Jacobian
for this transformation as
\begin{equation}
  \Biggl |
  \frac{\partial \bftheta_{k+1}}{\partial(\bftheta_k, u_1, u_2)}
  \Biggr | =
  \Biggl |
  \frac{\partial (w_{j_1}, w_{j_2}, \lambda_{j_1}, \lambda_{j_2})}
  {\partial (w_j, \lambda_{j}, u_1, u_2)}
  \Biggr | =
  \frac{\lambda_jw_j}{1-u_1} .
\end{equation}

For the reverse move, we select a pair of adjacent components $j_1$ and $j_2$.
Combining them, to get a new component labelled $j$, we set
\begin{equation*}
 w_j = w_{j_1} + w_{j_2},\quad
 \lambda_j = \frac{w_{j_1}\lambda_{j_1} + w_{j_2} \lambda_{j_2} }
 {w_{j_1} + w_{j_2}},
\end{equation*}
by keeping the first two moments of the proposed and current configuration
constant. We then sample a new set of allocation variables according to
Equation~\eqref{eq:allocconditional}.  We also keep track of the probability
of each allocation, so that $p_a(z)$ represents the probability of a given
allocation. To compute $p_a(z)$, we first simulate $z_i$ using
Equation~\eqref{eq:allocconditional}. For each $i$, the probability
of that allocation is given by
\begin{equation*}
  p_a(z_i) = \frac{w_{z_i} f(D_i | \lambda_{z_i}, E_i) }
  {\sum_{j=1}^k w_j f(D_i | \lambda_j, E_i)}.
\end{equation*}
Finally, we compute the probability of all allocations by
\begin{equation*}
  p_a(z) = \prod_{i=1}^n p_a(z_i).
\end{equation*}

\subsection{Acceptance Probability}
The acceptance probability of a move of type $(k, \bftheta_k) \Rightarrow (k',
\bftheta_{k'})$ is then $\min\{1,A_{k,k'}\}$, where
\begin{equation*}
  A_{k,k'} = \frac{\pi(k', \bftheta_{k'})}{\pi(k, \bftheta_k)} \times
  \frac{p(k'\Rightarrow k)}{p(k\Rightarrow k')} \times
  \frac{1}{q(u_1)q(u_2)} \times
  \Biggl| \frac{\partial \bftheta_{k'}}{\partial (\bftheta_k, u_1, u_2)}\Biggr|
\end{equation*}
\begin{multline*}
  A_{k,k'}=
  \frac{p(k')p(w'|\delta, k')p(\lambda'|\alpha,\beta,k')\bfL(D^n|\lambda',z')}
  {p(k) p(w|\delta, k) p(\lambda|\alpha,\beta, k) \bfL(D^n|\lambda, z) }
  \times \frac{p(z'| w', k+1) / p_a(z')}{p(z|w, k)/p_a(z)} \times \\
  \frac{p(k'\Rightarrow k)}{p(k\Rightarrow k')}
  \frac{1}{q(u_1)q(u_2)}
  \Bigl|\frac{\partial \bftheta_{k'}}{\partial (\bftheta_k, u_1, u_2)}\Bigr|  .
\end{multline*}
With $k'=k+1$ this becomes
\begin{multline*}
  A_{k,k+1}= \frac{p(k+1)}{p(k)}\times
  \frac{p(w'|\delta, k+1)}{p(w|\delta, k)}\times
  \frac{p(\lambda'|\alpha,\beta, k+1)}{p(\lambda|\alpha,\beta, k)}\times
  \frac{\bfL(D^n|\lambda', w')}{\bfL(D^n|\lambda, w)}\times \\
  \frac{p(k+1\Rightarrow k)}{p(k\Rightarrow k+1)}\times
  \frac1{q(u_1)q(u_2)}
  \Bigl|\frac{\partial \bftheta_{k+1}}{\partial(\bftheta_k, u_1, u_2)}\Bigr|.
\end{multline*}
Now with a uniform prior on the number of components $k$ and the weights $w$,
\begin{multline*}
  A_{k,k+1} = \frac{\Gamma(k+1)}{\Gamma(k)} \times
  \frac{(k+1) p(\lambda_{j_1}|\alpha, \beta) p(\lambda_{j_2}|\alpha, \beta)}
  {p(\lambda_j|\alpha,\beta)}
  \frac{p(z'| w', k+1) / p_a(z')}{p(z| w, k) / p_a(z)} \times \\
  \frac{\bfL(D^n | \lambda', z')}{\bfL(D^n| \lambda, z)} \times
  \frac{m_{k+1}}{s_k} \times
  \frac{1}{q(u_1) q(u_2)} \times
  \Bigl|\frac{\partial \bftheta_{k+1}}{\partial(\bftheta_k, u_1, u_2)}\Bigr|,
\end{multline*}
where the ratio of Gamma terms comes from the ratio of the prior
distributions on $w'$ and $w$ and
\begin{equation*}
  \frac{p(k+1\Rightarrow k)}{p(k\Rightarrow k+1)} =
  \frac{m_{k+1}/(k+1-1)}{s_k / k } =   \frac{m_{k+1}}{s_k } .
\end{equation*}

\subsection{Birth and Death Moves}\label{sec:birthdeath}
Suppose we are now at model $M_k$ with $k$ components, say
\begin{equation}\label{eq:oldparms}
  \bftheta_k =\{ (\lambda_1, w_1),
  \ldots, (\lambda_k, w_k) \}.
\end{equation}
If a move is proposed to increase the number of components by one, then we
simulate
\begin{equation*}
  \tilde {w} \sim Beta(1, k)
  \text{ and } \tilde
  \lambda \sim \mathcal{G}amma(a, b),
\end{equation*}
independently. The proposed new component will then have weight $\tilde w$
and the other weights are then scaled by a factor of $(1-\tilde w)$, so that
the sum of the weights remain $1$. The corresponding Poisson parameter for
the proposed component in $\tilde\lambda$.  Note that $\tilde\lambda$ is
sampled from its prior distribution. The proposed component is then
\begin{equation}\label{eq:newparms}
  \bftheta_{k+1} = ( \lambda_1, w_1/ (1-\tilde w)), \ldots, (\lambda_k,
  w_k/(1-\tilde w)), (\tilde \lambda , \tilde w) \} .
\end{equation}
Using this proposed value of $\bftheta_{k+1}$, we also simulate proposed
values for the allocations $z'$ with model $k+1$. Using the general form of
the reversible jump acceptance probability, see for example \citet{green1995}),
the probability of changing the number of components to $k+1$ is then
$\min\{1, A_{k,k+1}\}$, where
\begin{equation*}
  A_{k,k+1} = \frac{\pi(k+1, \bftheta_{k+1})}{\pi(k, \bftheta_k)} \times
  \frac{p(k+1\Rightarrow k)}{p(k\Rightarrow k+1)} \times
  \frac{1}{q(\tilde w)q(\tilde \lambda)} \times
  \Biggl|
  \frac{\partial \bftheta_{k+1}}{\partial (\bftheta_k,\tilde w,\tilde\lambda)}
  \Biggr|.
\end{equation*}
Making the necessary substitutions yield
\begin{multline}\label{eq:birthdeathaccept}
  A_{k,k+1} =
  \frac{
    p(k+1)p(w'|\delta,k+1)p(\lambda'|\alpha,\beta, k+1)\bfL(D^n|\lambda',z')}
  {p(k) p(w|\delta, k)  p(\lambda | \alpha, \beta, k) \bfL(D^n| \lambda, z)}
  \times\\
  \frac{p(z'|w', k+1)/p_a(z')}{p(z|w, k+1) / p_a(z)}\times
  \frac{p(k+1\Rightarrow k)}{p(k\Rightarrow k+1)} \times
  \frac{1}{q(\tilde w) q(\tilde \lambda)}
  \times
  \Bigl|
  \frac{\partial\bftheta_{k+1}}{\partial (\bftheta_k,\tilde w,\tilde\lambda)}
  \Bigr|.
\end{multline}
Using Equations~\eqref{eq:oldparms} and \eqref{eq:newparms} we then have the
Jacobian
\begin{equation*}
  \frac{\partial \bftheta_{k+1}}{ \partial (\bftheta_k,\tilde w,\tilde\lambda)}
  = (1-\tilde w)^{k-1}.
\end{equation*}
If we denote the probability of a birth when there are $k$ components by
$b_k$, and the probability of a death by $d_k$, with $b_k+d_k=1$, then
\begin{equation*}
  \frac{p(k+1\Rightarrow k) }{ p(k\Rightarrow k+1)}=
  \frac{d_{k+1}/ (k+1)}{ b_k },
\end{equation*}
since for the move to be reversible we would then be able to kill $k+1$
components in the new model, each with equal probability.  Substituting these
values in Equation~\eqref{eq:birthdeathaccept}, the ratio $A_{k,k+1}$
reduces to
\begin{multline*}
  A_{k,k+1} =
  \frac{\Gamma(k+1)}{\Gamma(k)} \times
  (k+1) p(\tilde\lambda) \times
  \frac{\bfL(D^n| \lambda',z')}{\bfL(D^n|\lambda,z)}
  \times
  \frac{p(z'|w'\ k+1)/p_a(z')}{p(z|w, k)/ p_a(z)}
  \times \\
  \frac{d_{k+1}/(k+1)}{b_k}
  \frac1{q(\tilde w) q(\tilde \lambda)}\times
  (1-\tilde w)^{k-1},
\end{multline*}
which on substituting $q(\tilde \lambda) = p(\tilde \lambda)$ and $q(\tilde
w) = k(1-\tilde w)^{k-1}$ further reduces to
\begin{equation}\label{eq:birthdeathaccept2}
  A_{k,k+1} =
  \frac{p(z'|w', k+1) / p_a(z')}{p(z|w, k)/ p_a(z)}
  \frac{\bfL(D^n| \lambda',z')}{\bfL(D^n|\lambda,z)}
  \times
   \frac{d_{k+1}}{b_k} .
\end{equation}
For a proposed death move, the acceptance probability is then
\begin{equation*}
  \min \{ 1, A_{k,k+1}^{-1} \}.
\end{equation*}
Even though the algorithm simulates new values of for the allocations when
proposing to move, it is not necessary to carry the allocations along. For
between--model moves, we could replace the missing data formulation by noting
that
\begin{equation*}
  \frac{p(z | w, k)}{p_a(z)}  \bfL(D^n | \lambda, z) =
  \bfL(D^n | \lambda, w).
\end{equation*}
Thus we could update model parameters using a scheme which does not require
conjugacy; see for example \citet{ctrj_mix,cappe2003}.

\section{Results}
We now present some numerical results for this dataset based on
the model described in Section~\ref{sec:mixformulation} and using the
algorithms described in Section~\ref{sec:heterrevjump}.

\begin{figure}[p]
  \subfigure[Birth/death and split/merge model trace.]{
    \label{fig:hetermod-bdsm}
    \begin{minipage}[b]{0.5\textwidth}
      \centering \includegraphics[width=\textwidth]{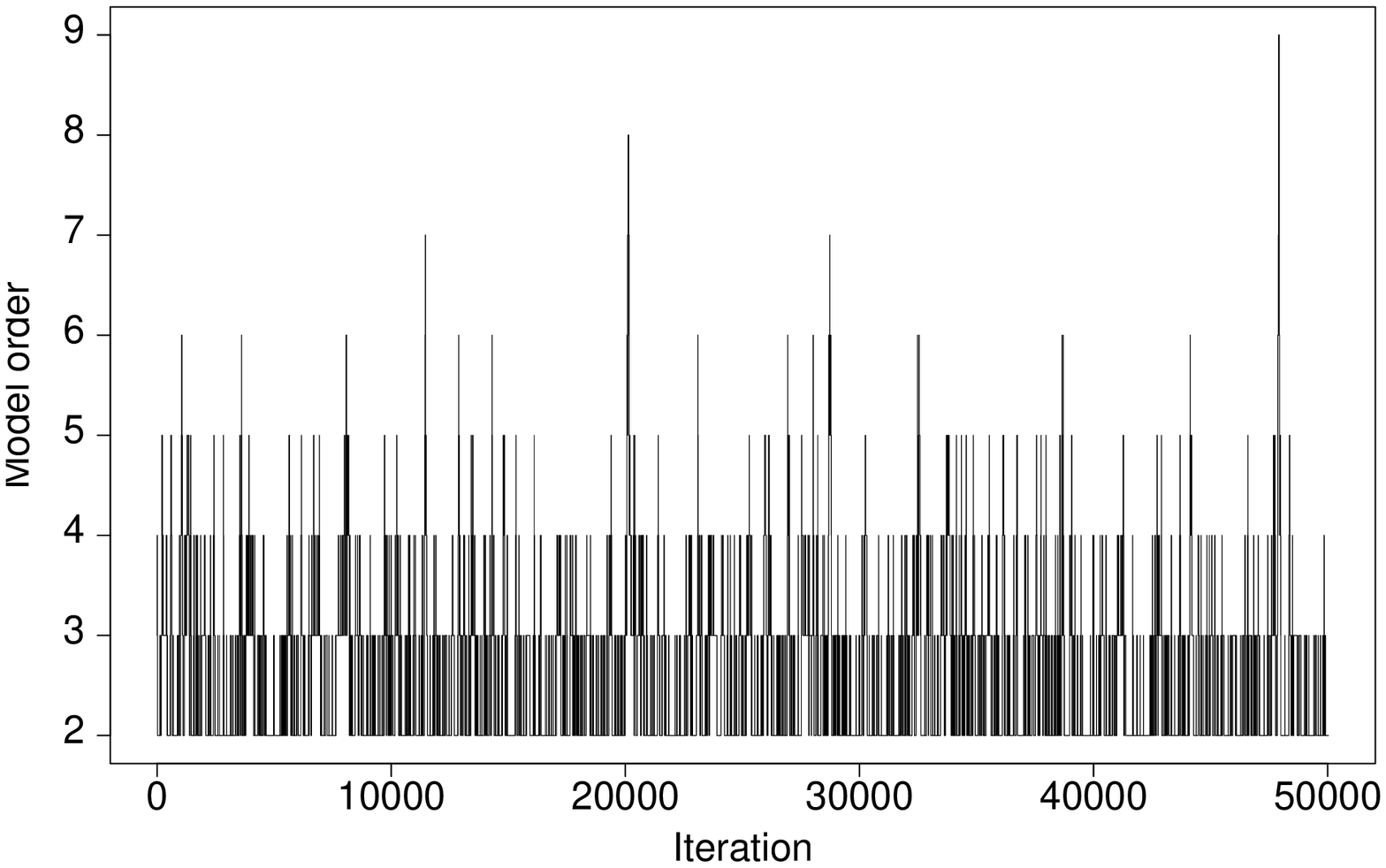}
    \end{minipage}}%
  \subfigure[Birth/death and split/merge histogram.]{
    \label{fig:heterhist-bdsm}
    \begin{minipage}[b]{0.5\textwidth}
      \centering
      \includegraphics[width=\textwidth]{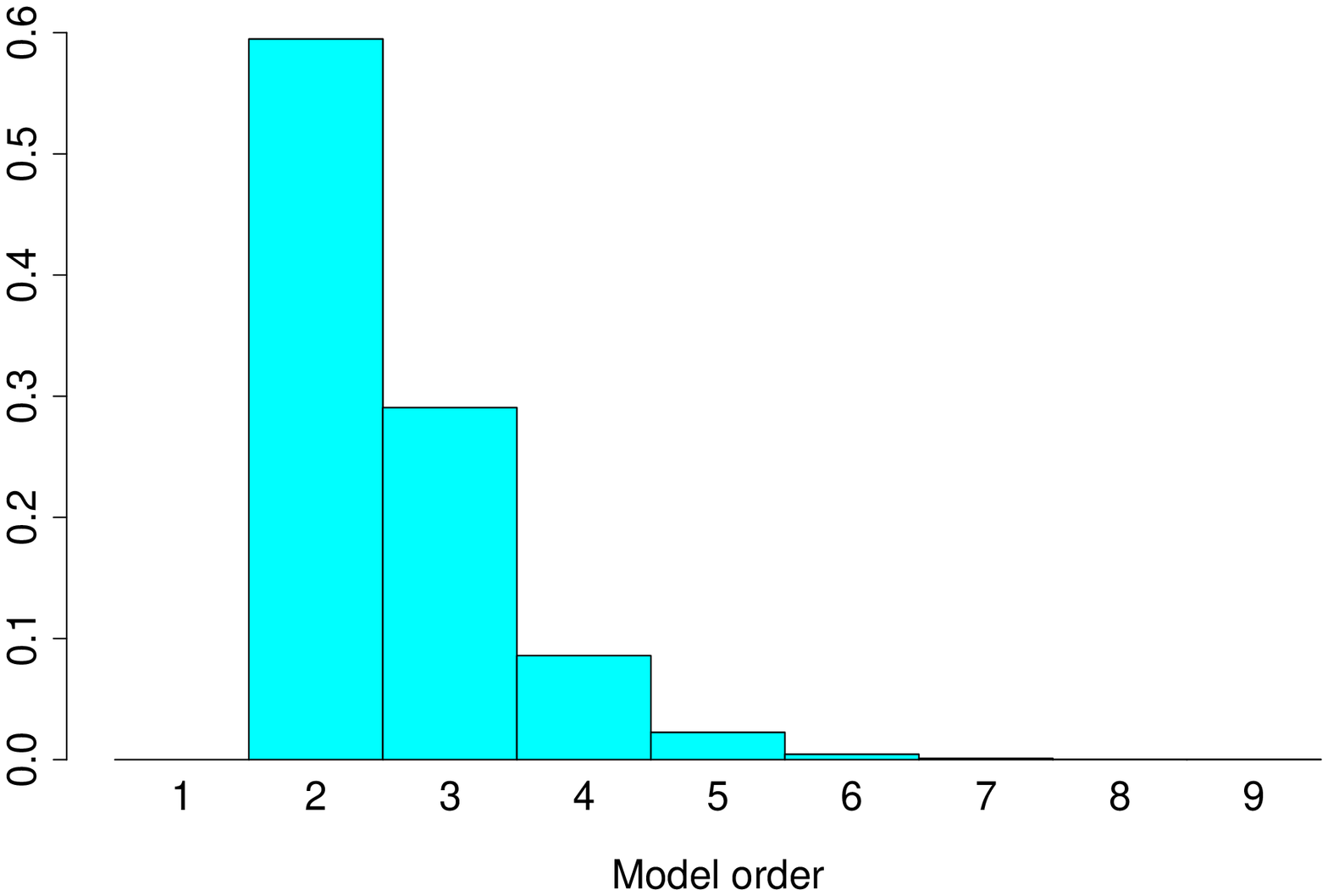}
    \end{minipage}} \\
  \subfigure[Birth/death model trace.]{
    \label{fig:hetermod-bd}
    \begin{minipage}[b]{0.5\textwidth}
      \centering \includegraphics[width=\textwidth]{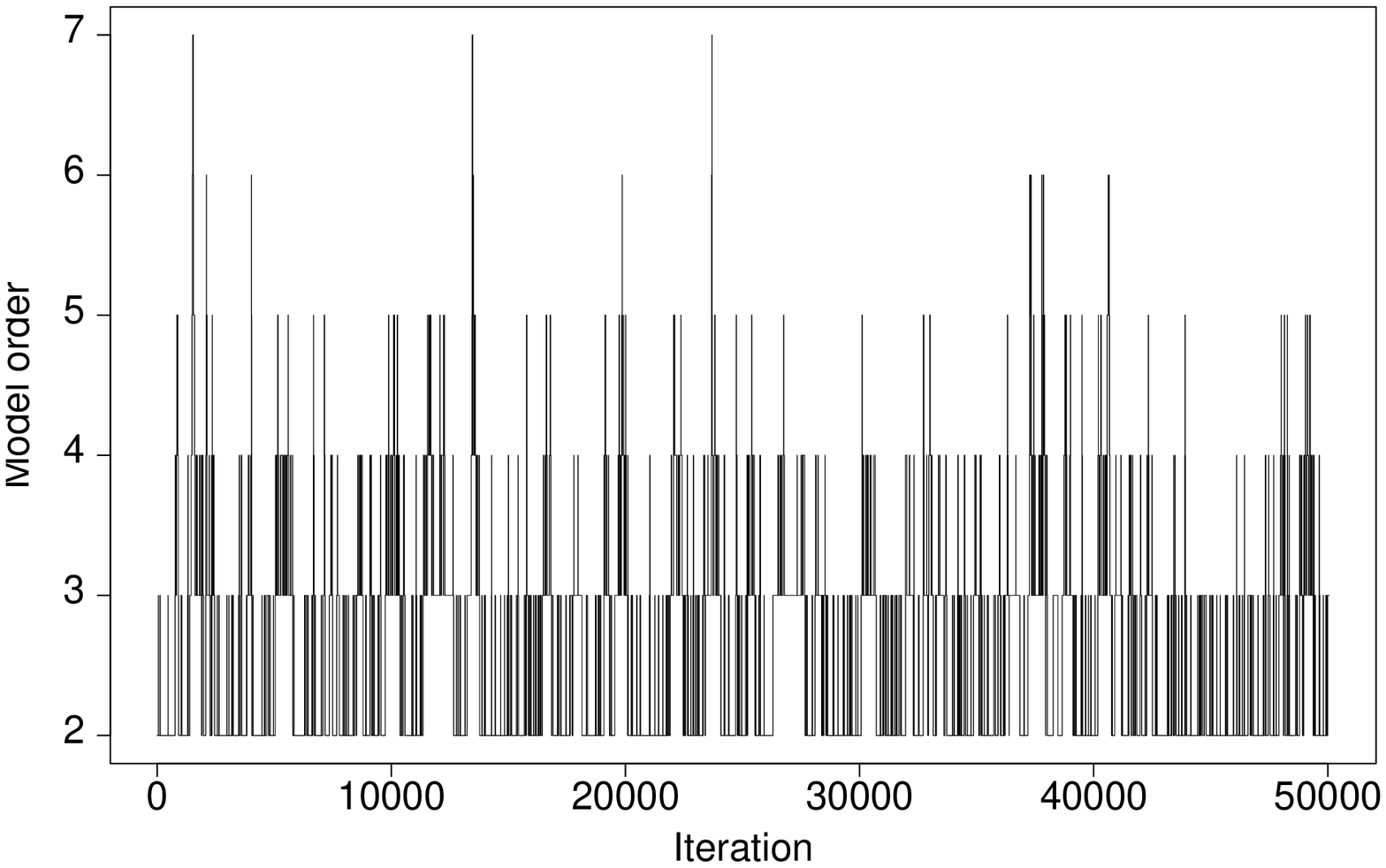}
    \end{minipage}}%
  \subfigure[Birth/death histogram.]{
    \label{fig:heterhist-bd}
    \begin{minipage}[b]{0.5\textwidth}
      \centering
      \includegraphics[width=\textwidth]{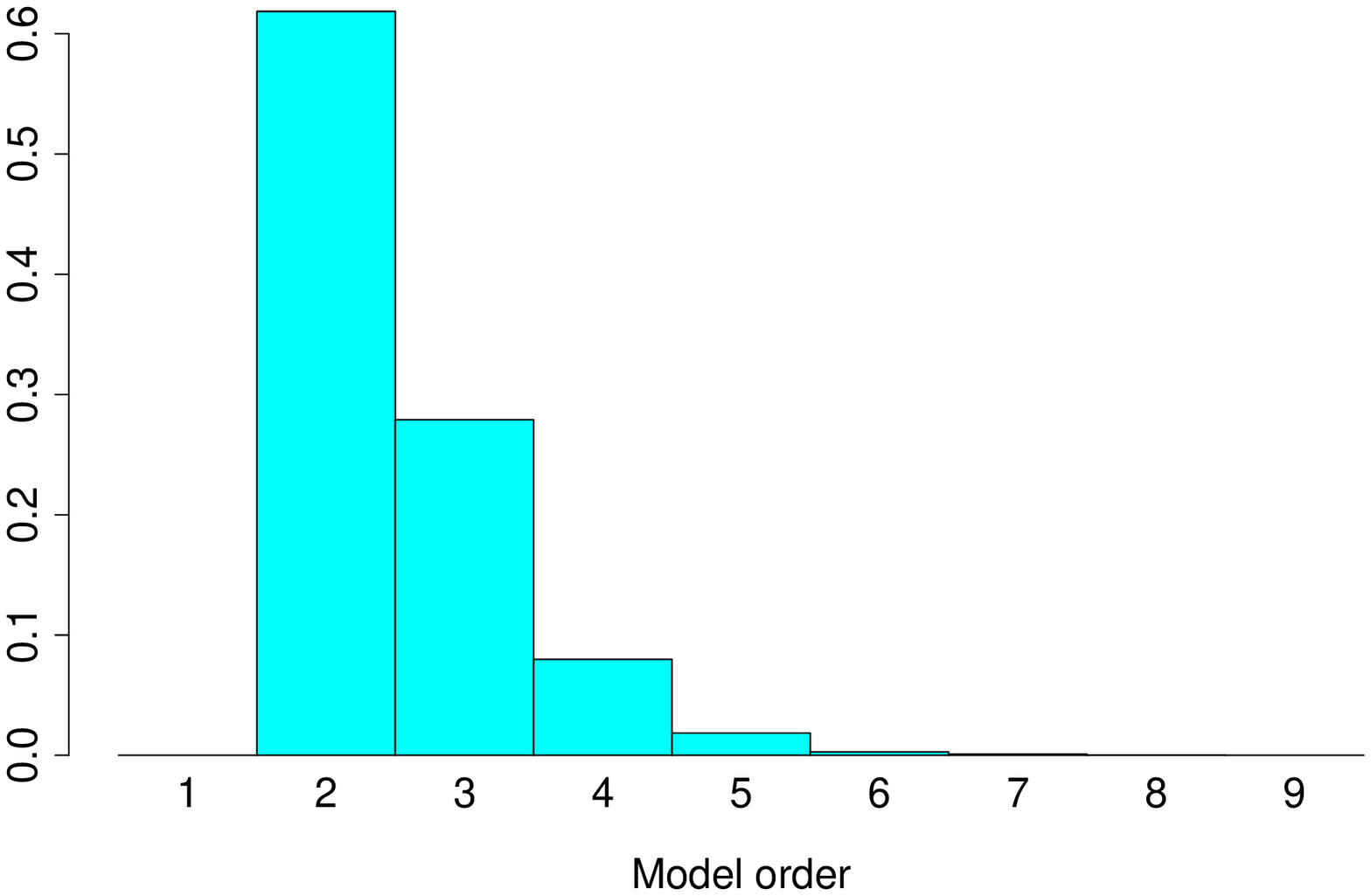}
    \end{minipage}}\\
  \subfigure[Split/merge model trace.]{
    \label{fig:hetermod-sm}
    \begin{minipage}[b]{0.5\textwidth}
      \centering \includegraphics[width=\textwidth]{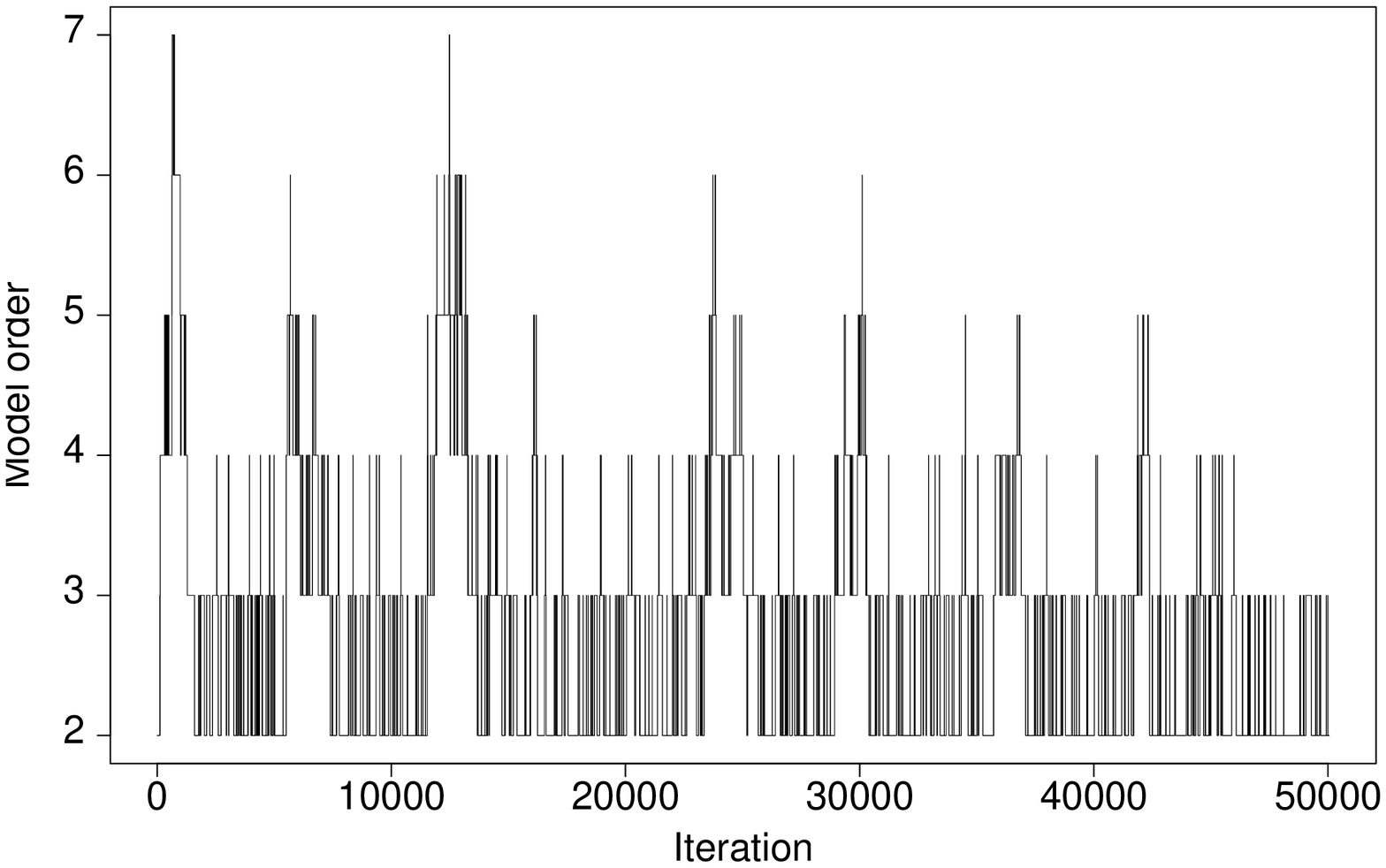}
    \end{minipage}}%
  \subfigure[Split/merge histogram.]{
    \label{fig:heterhist-sm}
    \begin{minipage}[b]{0.5\textwidth}
      \centering
      \includegraphics[width=\textwidth]{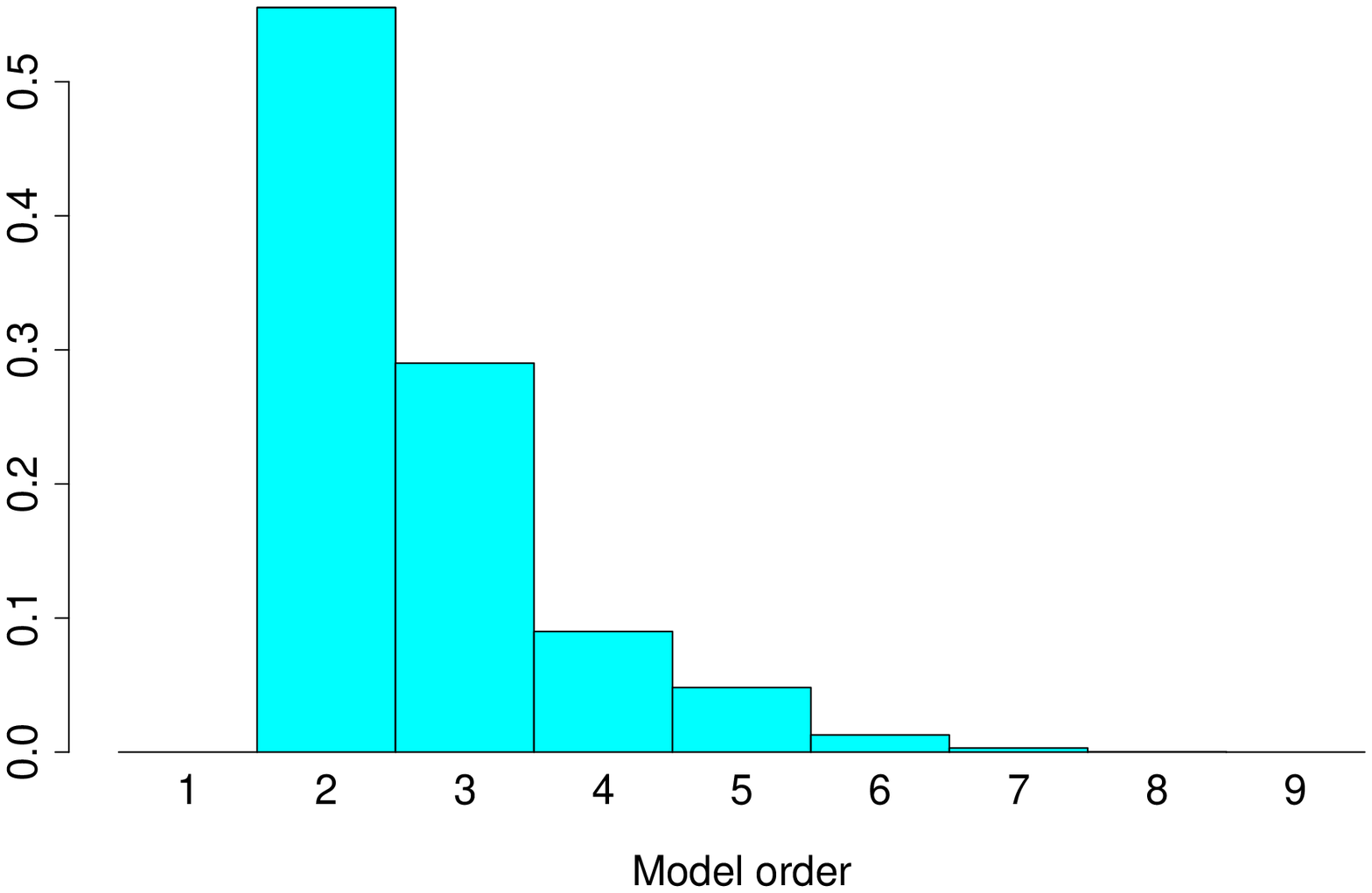}
    \end{minipage}}%
  \caption{Left: Model trace indicator.
    Right: Histogram of posterior model order.}
  \label{fig:hetermodhist-bdsm} 
\end{figure}

Table~\ref{tab:hetpost} shows the posterior model probabilities
calculated from the reversible jump algorithm by counting the
proportion of ties the algorithm visits each model.  A plot of the
number of components as the chain evolves is shown in
Figure~\ref{fig:hetermod-bdsm}. The results show clearly that the
number of components has a posterior mode at $k=2$. Also, the model
with $k=1$ component is never visited. If the algorithm is started
with $k=1$ then immediately it jumps to $k=2$ and never returns to
$k=1$. Since more than $88\%$ of the posterior probability mass is
placed on the models with $2$ or $3$ components, we discuss those
models in detail in Section~\ref{sec:hetermod2}. The between-- model
acceptance rates were $7.7\%$ and $5.5\%$ for the birth/death and
split/merge moves, respectively. The total acceptance rate when there
is equal probability of proposing a birth/death move or a split/merge
move, is $6.6\%$. These results are tabulated in
Table~\ref{tab:acceptrates}. 

\begin{figure}
    \centering
    \includegraphics[width=0.8\textwidth]{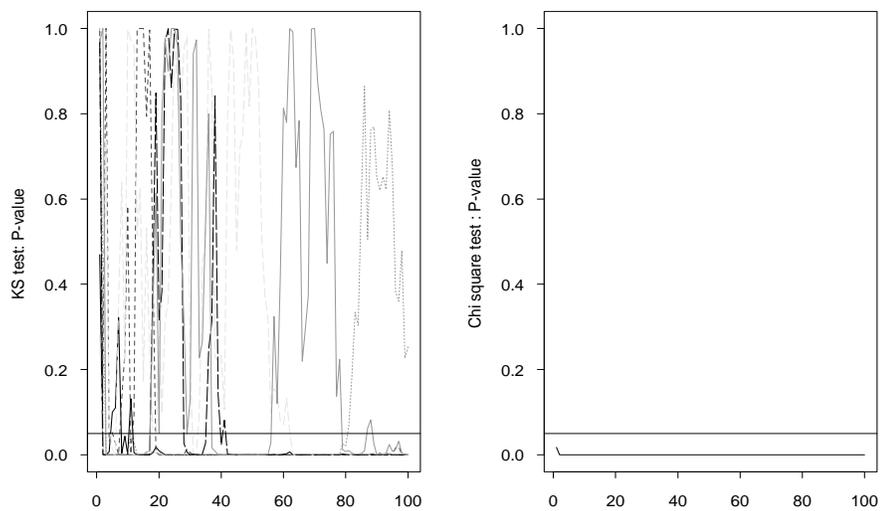}
    \caption{Convergence diagnostics.}\label{fig:het:conv}
\end{figure}

To assess convergence of the algorithm, we simulated 4
chains using different starting values and different random number
seeds for a total of 100000 iterations. Both the $\chi$--square and
Kolmogorov--Smirnov diagnostics are computed. These diagnostics are
plotted in Figure~\ref{fig:het:conv}.
\begin{table}
  \caption{\label{tab:hetpost}Posterior Model Order.}
  \centering
  \fbox{
    \begin{tabular}{r|c}
      \hline\hline
      Model Order &  Posterior Probability\\
      $k$ &  $\pi(k | D^n, E^n)$ \\
      \hline
      1 &0.00000\\
      2 &0.59485\\
      3 &0.29058\\
      4 &0.08588\\
      5 &0.02258\\
      6 &0.00448\\
      7 &0.00104\\
      8 &0.00034\\
      9 &0.00026\\
      10 &0.00000  \\
      \hline
    \end{tabular}}
\end{table}
\begin{table}
  \caption{\label{tab:acceptrates}Acceptance Rates.}
  \centering
\fbox{
  \begin{tabular}{cc}
    \hline\hline
    Scheme & Acceptance Rate \\
    \hline
    Birth/Death & 0.077\\
    Split/Merge & 0.055\\
    Birth/Death and Split/Merge& 0.066\\
    \hline
  \end{tabular}}
\end{table}

\subsection{Comparing the Model Move Schemes}
A comparison of the individual acceptance probabilities shows that the between
model moves are accepted with a larger rate for the birth death scheme
compared with the split merge scheme. This might not always be true, as
other split merge schemes may be proposed \citep{viallefont2002}. It is
interesting to note that although the birth and death rates are higher than
the split and merge rates, the combined scheme seems to mix better than either
scheme implemented alone. Even though the birth and death scheme have a
higher acceptance rate for between--model moves, the excursions away from the
values of highest posterior density, $k=2$ and $k=3$, are longer than for the
combined scheme or the split and merge scheme since. This is because when
proposing parameters independently from the prior, areas of low probability
can be proposed, whereas, with the split and merge scheme, areas of low
probability mass will generally be rejected.  Based on the results presented here,
 the birth/death method would be the preferred algorithm.

\subsection{Detailed Results for $k=2$ and $k=3$}\label{sec:hetermod2}

A histogram plot of the model indicator from the reversible jump algorithm of
Section~\ref{sec:heterrevjump} shows that the most plausible model generating
the claims in the portfolio is a mixture of two Poisson distributions. In
this section we look further at the results, conditional on their being only
two components, or three components, in the mixture.

Recall the missing data formulation introduced in
Section~\ref{sec:mixformulation} for the number of components
conditional on $k=2$, we observed the posterior distribution of $z$ at
each iteration when $k=2$. A study of values of $z$ will tell us how
the data has been allocated to the components and therefore, which
data points have been generated from either the first Poisson
distribution or the second Poisson distribution. This information,
along with further information from the portfolio, will help insurance
companies classify groups of life insurance portfolios.  
The parameter estimates are shown in Table~\ref{tab:hetermod2}.


\begin{table}
  \caption{  \label{tab:hetermod2}Parameters Estimates Conditional on $k=2$.}
  \centering
  \fbox{
  \begin{tabular}{r|rr}
    \hline \hline
    &Estimate & 95\% HPD Interval \\
    \hline
    $\lambda_1$ & 0.731 & (0.626, 0.839)  \\
    $w_1$       & 0.636 & (0.428, 0.821) \\
    $\lambda_2$ & 1.896 & (1.557, 2.249)  \\
    $w_2$       & 0.363 & (0.178, 0.571) \\
    \hline
  \end{tabular}}
\end{table}
\begin{table}
  \caption{\label{tab:hetermod3}Parameters Estimates Conditional on $k=3$.}
  \centering
\fbox{
  \begin{tabular}{r|rr}
    \hline \hline
    &Estimate & 95\% HPD Interval \\
    \hline
    $\lambda_1$ & 0.462 & (0, 0.770) \\
    $w_1$       & 0.299 & (0.000, 0.656) \\
    $\lambda_2$ & 1.115 & (0.625, 1.692) \\
    $w_2$       & 0.495 & (0.157, 0.797) \\
    $\lambda_3$ & 2.481 & (1.570, 3.366) \\
    $w_3$       & 0.204 & (0.002, 0.464) \\
    \hline
  \end{tabular}}
\end{table}

Similar results for the posterior parameter estimates, conditional on
there being three components in the mixture, are given in
Table~\ref{tab:hetermod3}.

\begin{figure}
  \subfigure[$k=2$]{
    \label{fig:hetalloc2}
    \begin{minipage}[b]{0.5\textwidth}
      \centering
      \includegraphics[width=\textwidth]{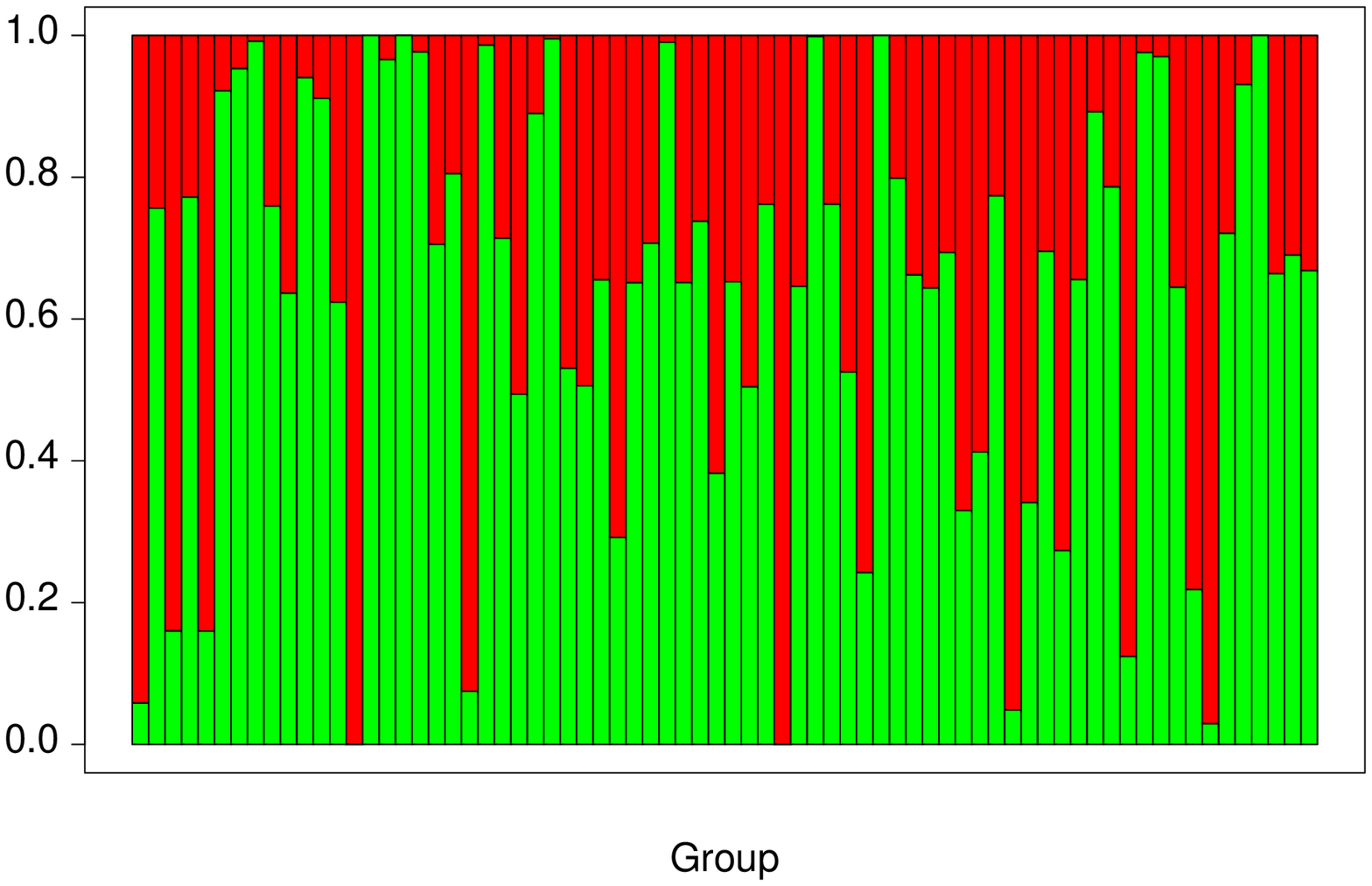}
    \end{minipage}}%
  \subfigure[$k=3$]{
    \label{fig:hetalloc3}
    \begin{minipage}[b]{0.5\textwidth}
      \centering
      \includegraphics[width=\textwidth]{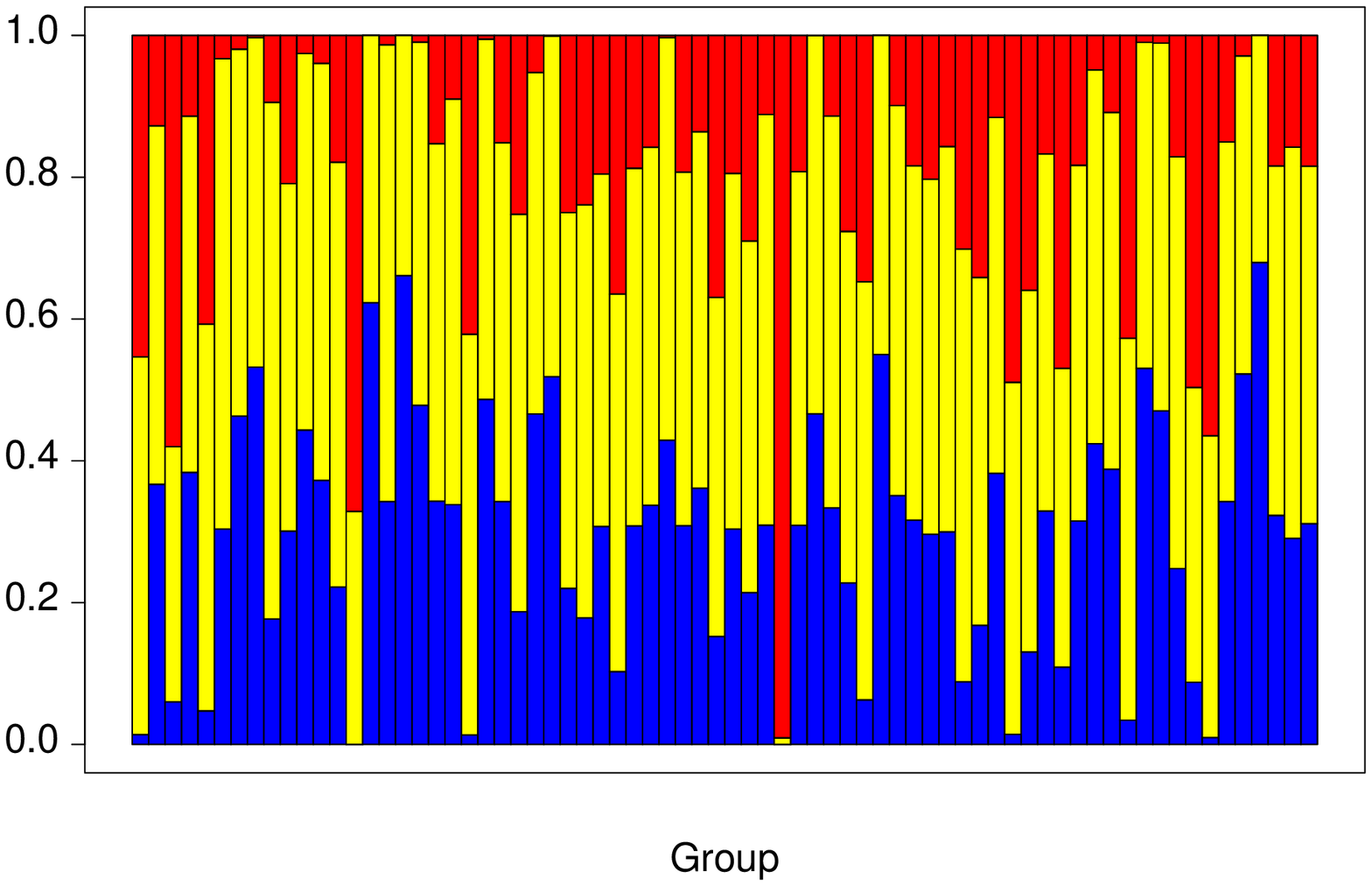}
    \end{minipage}}%
  \caption{Probability (vertical axis) of data from Group $i$ (horizontal axis)
    being assigned to individual components conditional on $k=2$ (left) and
  $k=3$ (right).}
  \label{fig:hetalloc}
\end{figure}

Figure~\ref{fig:hetalloc} show the posterior probability of each data
point being allocated to a particular component of the mixture,
conditional on $k=2$ and conditional $k=3$, respectively.

\section{Summary}
We present a model for heterogeneity in group life insurance. We show
that the assumption of identical heterogeneity for all groups under
consideration, may not necessarily hold. In this case, it is necessary
to put similar groups together for further analysis.  We employ a
non--parametric approach any apply reversible jump methods to
determine the number of components in the mixture. An extension of the
current work to the case where claims are grouped, such as
\citep{walhin1999, walhin2000}, would therefore be appropriate.

\clearpage
\bibliographystyle{chicago}
\bibliography{gbthesis}

\end{document}